# Orthogonal Attosecond Control of Solid-State Harmonics by Optical Waveforms and Quantum Geometry Engineering

Zhenjiang Zhao[1], Zhihua Zheng[1], Zhiyi Xu[1], Xing Ran[1], Xiaolong Yao[1,3,*], Fangping Ouyang[1,2,†]

[1]*School of Physical Science and Technology, Xinjiang Key Laboratory of Solid-State Physics and Devices, Xinjiang University, Urumqi 830017, China*

[2]*School of Physics, Institute of Quantum Physics, Hunan Key Laboratory for Super-Microstructure and Ultrafast Process, and Hunan Key Laboratory of Nanophotonics and Devices, Central South University, Changsha, 410083, China*

[3]*Beijing Computational Science Research Center, 100193 Beijing, China*

[*]Contact author: xlyao@xju.edu.cn

[†]Contact author: ouyangfp06@tsinghua.org.cn

## Abstract

High-harmonic generation (HHG) in two-dimensional materials offers a compelling route toward compact extreme ultraviolet sources and probing electron dynamics on the attosecond scale. However, achieving precise control over the emission and disentangling the complex interplay between intraband and interband quantum pathways remains a central challenge. Here, we demonstrate through first-principles simulations that HHG in monolayer $WS_2$ can be subjected to precise, complementary control by combining all-optical two-color laser fields with mechanical strain engineering. This dual-mode strategy provides distinct, orthogonal control over harmonic yield, polarization, and spectral features. We reveal that sculpting the two-color field's relative phase provides a sub-femtosecond switch for the quantum coherence of electron-hole pairs, thereby optimizing harmonic emission. Crucially, we uncover that tensile strain modulates the total harmonic yield and specifically amplifies the perpendicular harmonic component by nearly a factor of two. This enhancement arises through a dual mechanism - while strain-modified band dispersion enhances the intraband current, a significant reshaping of the Berry curvature (BC) substantially increases the anomalous velocity contribution to the interband response. This quantum geometric effect manifests as a robust, monotonic dependence of the harmonic yield on strain and a significant amplification of the perpendicularly polarized harmonics, providing a clear experimental signature for probing quantum geometric effects. Our findings establish a versatile framework for optimizing solid-state HHG and introduce a powerful all-optical method to map strain and quantum geometric properties of materials, positioning monolayer $WS_2$ as a model system for exploring attosecond physics at the nexus of bulk and atomic scales.

**Keywords:** high-harmonic generation, strain engineering, two-color laser, attosecond dynamics, Berry curvature

## 1. Introduction

High-harmonic generation (HHG), a quintessential nonlinear optical process, arises from the interaction of intense laser fields with matter.[1] This phenomenon has garnered significant interest across atomic, molecular, and solid-state systems.[2-9] In solids, the high electron density and lattice periodicity impose complex constraints on electronic structure and momentum, imparting distinct characteristics to the HHG mechanism compared to its gas-phase counterpart.[10-13] These unique solid-state features have spurred the development of compact and efficient high-harmonic sources and hold significant promise for applications in emerging fields such as multi-terahertz electronics and all-optical information processing.[14-16]

Among condensed matter systems, two-dimensional (2D) materials - such as graphene, hexagonal boron nitride, and transition metal dichalcogenides (TMDs) - have emerged as ideal platforms for studying ultrafast electronic dynamics due to their unique symmetries and electronic properties.[5, 9, 17-19] These materials exhibit a remarkable duality, bridging atomic- and bulk-like responses. For instance, in monolayer hexagonal boron nitride (hBN), when the driving laser field is polarized perpendicular to the material plane, the real-space electron trajectories resemble those in isolated atoms.[9, 20] Conversely, when the laser is polarized in-plane, the response is characteristic of bulk HHG. This duality positions 2D materials as a critical bridge between atomic-scale and condensed matter physics.

Precise control over HHG has been demonstrated by tailoring laser field parameters (e.g., polarization, ellipticity, and waveform) or by modifying material properties (e.g., applying strain , defect engineering or varying layer numbers).[17, 21-23] Complementary to these static approaches, recent studies have also established that coherent phonon dynamics can induce dynamic symmetry breaking, offering an effective avenue for modulating harmonic generation[24]. Specifically, two-color laser fields, where the relative phase between the fundamental and second harmonic is controlled, enable sub-femtosecond manipulation of electron trajectories, leading to modulated harmonic yields and extended cutoff energies. Strain engineering, a powerful tool for tuning material properties, can modify the band structure and break structural symmetries.[25-27] For example, it has been shown that strain-induced modifications to the band structure and Berry curvature (BC) can enhance both intraband and interband currents, thereby boosting HHG.[8]

Solid-state HHG originates from the coherent interference of two fundamental charge transport mechanisms - interband transitions and intraband acceleration. The total macroscopic current density, $\boldsymbol{J}(t)$, which dictates the far-field harmonic emission, is governed by the dynamic interplay of these complementary processes[28]. Disentangling their respective contributions is essential for controlling sub-optical-cycle electron dynamics in 2D materials and for engineering tailored attosecond sources. Specifically, intraband currents stem from the coherent acceleration of Bloch wavepackets within individual energy bands, driven by the strong-field vector potential. Semiclassically, the wavepacket velocity $\boldsymbol{V}_n(\boldsymbol{k},t)$ encompasses a conventional group velocity determined by the band dispersion, $\partial_{\boldsymbol{k}}\varepsilon_{n\boldsymbol{k}}$, and an anomalous transverse velocity, $\boldsymbol{E}(t)\times\boldsymbol{\Omega}_n(\boldsymbol{k})$, dictated by the BC $\boldsymbol{\Omega}_n(\boldsymbol{k})$ [29, 30]. Acting as an intrinsic momentum-space pseudo-magnetic field, the BC fundamentally steers the sub-cycle trajectories of charge carriers - an effect heavily pronounced in lattices with broken inversion symmetry[29, 30]. Crucially, this anomalous velocity drives transverse charge currents, predominantly sourcing the generation of perpendicularly polarized harmonics. This orthogonal emission serves as a direct optical signature of the material's quantum geometry, particularly in systems where structural symmetries are dynamically or statically broken by external perturbations such as mechanical strain [29].

On the contrary, interband current originates from dipole-allowed transitions of electrons between different energy bands, such as from the valence band to the conduction band. As elucidated by detailed electron dynamics analyses based on the semiclassical model[31, 32], this process is fundamentally quantum mechanical, undergoing a complete cycle of electron-hole pair generation (ionization), subsequent intraband acceleration driven by the laser field, and final coherent recombination. The amplitude and phase information of these coherent interband transitions are encoded in the off-diagonal elements of the density matrix, ultimately leading to the emission of high-harmonic photons through the direct recombination of electron-hole pairs. When the driving field is linearly polarized along a high-symmetry crystal axis (e.g., the $x$-axis), this interband mechanism predominantly yields harmonics polarized parallel to the applied field. This collinear emission assumes that the local band dispersion along the polarization axis is weakly anisotropic. Conversely, breaking this dynamic symmetry - either by aligning the driving field along a low-symmetry direction or through pronounced intrinsic band anisotropy and complex multi-band couplings - drives transverse electron dynamics, thereby generating harmonics with highly complex polarization states. In summary, under typical conditions (e.g., linear driving polarized along a high-symmetry crystal axis), the total HHG signal results from the coherent superposition of intraband and interband currents. The interband current primarily contributes to harmonics polarized parallel to the driving field (e.g., the $P_x$ component), with its physical picture corresponding to the generation and direct recombination radiation of electron-hole pairs. In contrast, the intraband current, particularly the anomalous velocity term mediated by the BC, is the key physical origin for generating harmonic components polarized perpendicular to the driving field (e.g., the $P_y$ component). This is especially prominent in systems with broken time-reversal symmetry. Conversely, intraband current arises from the field-driven coherent acceleration of Bloch wavepackets within individual energy bands [33]. The dynamical competition and relative dominance of these intraband and interband pathways are rigorously dictated by the material's local band dispersion and underlying quantum geometry. This sensitivity is starkly illustrated in flat-band systems such as $NbOCl_2$; in these environments, the negligible band dispersion strongly quenches the intraband velocity, leaving interband polarization and subsequent recombination to exclusively govern the high-harmonic emission [34].

Despite this progress, the interplay between intraband and interband contributions to solid-state HHG remains a subject of active debate.[2, 11, 13, 15, 35-37] The relative dominance of these two mechanisms varies significantly with the material system and spectral range, accounting for the diversity of experimental observations. Therefore, a deeper understanding of the factors governing strong-field electron dynamics in 2D materials is crucial. Such knowledge is essential for developing effective control strategies and advancing both attosecond science and condensed matter physics.

In this work, we employ first-principles simulations based on real-time time-dependent density functional theory (rt-TDDFT),[38, 39] as implemented in the octopus code,[40] to systematically investigate the complementary effects of two-color phase modulation and strain engineering on the sub-femtosecond electron dynamics governing HHG in monolayer $WS_2$ (1L-$WS_2$). While strain engineering has been explored in other TMDs such as $MoS_2$[8], where compressive strain was reported to enhance HHG driven by intraband dynamics, we reveal a fundamentally distinct regime in monolayer $WS_2$. Here, we demonstrate that tensile strain acts as the primary amplifier for the perpendicularly polarized harmonics, which are governed primarily by intraband dynamics (anomalous velocity). Crucially, by applying strain engineering, we elucidate how modifications to the band structure and BC influence the polarization-resolved harmonic yields, providing a sensitive probe of band topology. Furthermore, we introduce an orthogonal control dimension - two-color phase modulation - to actively manipulate the quantum coherence of these transitions, a degree of freedom not addressed in prior static strain studies. This work provides not only an experimentally feasible strategy for controlling HHG in 2D materials but also new theoretical insights into the quantum dynamics of laser-solid interactions.

## 2. Theoretical and computational methodology

Electronic dynamics were simulated using real-time time-dependent density functional theory (rt-TDDFT) as implemented in the Octopus code.[40] We employed the adiabatic local density approximation (ALDA) with norm-conserving HGH pseudopotentials.[41] To accurately capture high-energy conduction states, the TDKS equations were discretized on a 0.21 Å real-space grid (effective cutoff ~850 eV). We acknowledge that under strong-field driving, HHG in TMDs is a complex process involving highly delocalized electron dynamics across the entire Brillouin zone, including contributions and quantum interferences from the Γ and M points, as recently demonstrated by Kim *et al*. [42]. By employing a dense 36×36×1 Monkhorst-Pack grid over the full Brillouin zone, our rt-TDDFT simulations inherently capture these comprehensive BZ-wide dynamics in the total current $\boldsymbol{J}(t)$. However, the central focus of this study is the orthogonal control of the perpendicularly polarized harmonics ($P_y$). Unlike the parallel interband transitions that span the BZ, the perpendicular harmonic generation is rigorously governed by the anomalous velocity term, which is driven by the BC. Because the BC in monolayer $WS_2$ is overwhelmingly concentrated at the direct bandgap K and K' valleys due to strong spin-orbit coupling and broken inversion symmetry, the physical origin of the orthogonal HHG modulation remains deeply rooted in the quantum geometry of these high-symmetry valleys. Thus, while our computational framework encompasses the full BZ, our analytical focus on the K/K' valleys serves to isolate and explain the specific topological signatures encoded in the perpendicular harmonic yield.

The system is driven by a two-color laser field $\boldsymbol{E}(t)$, linearly polarized along the zigzag direction of the monolayer $WS_2$ [see Figure 1(a)]. The field is a superposition of a fundamental pulse and its second harmonic, derived from the total vector potential $\boldsymbol{A}(t)$, where $\boldsymbol{E}(t) = -\frac{\partial \boldsymbol{A}(t)}{\partial t}$

$$\boldsymbol{A}(t) = f(t)[\boldsymbol{A}_0 cos\,(\omega_0 t) + \boldsymbol{A}_1 cos\,(\omega_1 t + \Delta\phi)] \qquad (1)$$

Both components share a common Gaussian envelope $f(t)$

$$f(t) = exp\,(-\frac{(t-t_0)^2}{2\tau^2}) \qquad (2)$$

In these expressions, $\boldsymbol{A}_0(\omega_0)$ and $\boldsymbol{A}_1(\omega_1=2\omega_0)$ are the vector potential amplitudes (frequencies) of the fundamental and second-harmonic fields, respectively. For all simulations, the fundamental photon energy is set to $\hbar\omega_0 = 0.517$ eV ($\lambda_0$=2400 nm). The Gaussian pulse envelope is defined by a standard deviation of $\tau$=12 fs and is centered at $t_0$=50 fs. This corresponds to a $1/e$ intensity duration of $2\tau$=24 fs and an intensity full width at half maximum (FWHM) of $2\sqrt{2ln2}\cdot\tau \approx 28.3\,fs$. The intensity of the fundamental field is fixed at $I_0$=$10^{12}$ W/cm$^2$. The second-harmonic intensity, controlled via $\boldsymbol{A}_1$, and the relative phase, $\Delta\phi$, serve as the key control parameters in our study.

**3. Results and discussion**

Figure 1(a) illustrates the crystal lattice of 1L-$WS_2$ and schematically depicts the HHG process under a two-color laser field. The electronic band structure, calculated from first principles, is shown in Figure 1(b). Our calculations, employing the local density approximation (LDA), yield a minimum direct bandgap of 1.7 eV at the K/K' points. While this calculated bandgap is indeed smaller than the experimental value of 2.0 eV, a well-known and systematic limitation of LDA functionals in semiconductors, this underestimation primarily impacts the quantitative predictive power of certain metrics, such as the exact harmonic cutoff energy. Nevertheless, the LDA functional accurately reproduces the essential qualitative dispersion characteristics of the valence and conduction bands and the underlying wavefunction topology, which are critical for modeling electron dynamics and are validated for 2D materials.[43] However, our key qualitative conclusions concerning the control of carrier dynamics, the effects of symmetry breaking, and the crucial role of the BC are robust, as these phenomena are primarily dictated by the band dispersion and wavefunction topology, which are well-described by the LDA functional. Future work employing more advanced methods, such as hybrid functionals or GW-corrected band structures, could refine the absolute cutoff energies; however, these quantitative adjustments would not alter our core mechanistic insights related to topology and symmetry. For reference, the Perdew-Burke-Ernzerhof (PBE) functional predicts a slightly larger bandgap, consistent with prior theoretical work.[44] The reciprocal space structure is depicted in Figure 1(c), with a calculated lattice constant of $a$=3.186 Å.

To probe the nonlinear optical response, we investigate the influence of the relative phase, $\Delta\varphi$, between the two components of a linearly polarized two-color laser field on HHG in 1L-$WS_2$. This parameter is known to be a powerful tool for controlling harmonic emission.[9, 17, 33, 45, 46] As illustrated by the electric field waveforms in Figure 2(a), the two-color scheme introduces a pronounced temporal asymmetry controlled by $\Delta\varphi$. While the fundamental pulse ($\beta$=0) is symmetric (black curve), the synthesized two-color fields ($\beta$=1) exhibit strong asymmetry for both $\Delta\varphi$=0.7π (red) and $\Delta\varphi=\pi$ (blue). It is clear that the participation of the frequency-doubled pulses leads to a significant extension of the cutoff photon energy, together with an enhancement of harmonic yields. Meanwhile, the shape of the spectra is sensitive to the relative phase between the two components [Figure 2(b)].

This broken symmetry has profound consequences for the harmonic polarization. As shown in Figures 2(c) and 2(d), the two-color field simultaneously generates both odd- and even-order harmonics and allows for the dynamic redistribution of harmonic energy between the parallel ($P_x$, along the $x$-direction) and perpendicular ($P_y$, along the $y$-direction) polarization components. By performing a polarization decomposition of the harmonic signal, we can disentangle the microscopic origins of the emission. This method can reveal the specific distribution of harmonic intensities across different polarization directions, thereby more clearly illustrating the contributions of distinct mechanisms. While fully distinguishing in-band and inter-band contributions may prove complex, explicitly presenting the intensity spectra of the $P_x$ and $P_y$ components aids in understanding their relative importance within the higher-order harmonic signal. In the nonlinear optical response of light-matter coupling, interband transitions act as the key mechanism for inducing intraband currents by coherently altering the population distribution between the conduction and valence bands. Photon-driven interband transitions selectively generate non-equilibrium electron-hole populations on specific crystal momentum states, disrupting the system's original momentum equilibrium. This non-equilibrium population distribution directly confers a net group velocity shift to electrons within the band structure, thereby contributing a significant intraband current. This process can be explicitly described as "intraband current induced by interband transitions"[47], where the rearrangement of population serves as an intermediate physical quantity, effectively bridging the interband coupling of the light field with the intraband transport response of the system. The parallel component ($P_x$) is overwhelmingly dominant, contributing over 90% of the total intensity, and originates almost exclusively from interband currents. In contrast, the much weaker perpendicular component ($P_y$) arises primarily from intraband dynamics - the acceleration of charge carriers within a single band - and exhibits a more sensitive dependence on $\Delta\varphi$[8]. To rigorously quantify these mechanisms, we performed time-dependent orbital projection calculations (see Figures 2(e), (2f) and Supplementary Figures S1-S10). Our numerical analysis reveals that for the parallel polarization ($P_x$), interband transitions account for over 90% of the total harmonic yield. Conversely, the perpendicular component ($P_y$), while weaker, shows a distinct dependence on intraband currents derived from the anomalous velocity. This analysis unequivocally demonstrates that interband transitions are the dominant mechanism governing the total HHG yield in 1L-$WS_2$, while intraband dynamics are crucial for the orthogonal polarization control.

The relative phase $\Delta\varphi$ is therefore a key parameter for controlling not just the polarization but also the overall efficiency of the HHG process. Figure 3(a) quantifies this dependence by plotting the integrated harmonic yield across different spectral bands as a function of $\Delta\varphi$. The total yield (orders 1-40) and the high-energy plateau yield (orders 4 - 30) exhibit a clear modulation, reaching a maximum at $\Delta\varphi\approx0.7\pi$ and a minimum near $\Delta\varphi$=0 and π. Quantitatively, as shown in Figure 3(a), varying $\Delta\varphi$ modulates the total yield by approximately 10%. While modest in magnitude compared to strain effects, this phase dependence provides a critical knob for optimizing the high-energy plateau (~22% modulation) and controlling the cutoff energy. The modulation of HHG efficiency is governed by the sub-cycle interference of the two-color field. As shown in the time-frequency analysis (Figure 4), $\Delta\varphi$ controls the constructive superposition of the fundamental and second-harmonic fields. Specifically, $\Delta\varphi$=0.7π maximizes the instantaneous field amplitude at the moment of recombination, whereas $\Delta\varphi$=π leads to destructive interference in the waveform, suppressing the peak effective field and the resulting harmonic cutoff.

To further explore independent control dimensions, we investigated another direct physical parameter - laser intensity. Under fixed phase relationships, simply increasing the intensity of the driving laser not only enhances the overall yield across the entire harmonic spectrum but also broadens the plateau region and increases the cutoff energy for higher-order harmonics [Figure 3(b)]. This confirms that laser intensity serves as another core means for regulating HHG yield and spectral characteristics. Its effects - yield enhancement and plateau broadening - exhibit distinct physical mechanisms and outcomes compared to pure phase modulation.

To further elucidate the microscopic dynamics underlying the harmonic emission, we performed a wavelet time-frequency analysis of the induced electronic current. This technique maps the temporal evolution of the current, $\boldsymbol{j}(t)$, into a time-frequency domain, $S(t, \omega)$, providing a time-resolved picture of the emission process[48]

$$S(t_0, \omega_0) = \int j(t) w^*(t - t_0, \omega_0) dt \quad (3)$$

Here, $\boldsymbol{j}(t)$ is the microscopic current and $w^*(t, \omega)$ is the complex conjugate of the mother wavelet. For this analysis, we employ a Gabor wavelet, which provides optimal resolution in both time and frequency

$$w(t, \omega_0) = \left(\frac{1}{\sigma_t\sqrt{2\pi}}\right) exp\left(-\frac{t^2}{2\sigma_t^2}\right) exp(i\omega_0 t) \quad (4)$$

where $\sigma_t$ determines the temporal duration of the wavelet window. The resulting time-frequency intensity distribution, which is what is plotted, is given by the squared modulus of the transform

$$I(t_0, \omega_0) = | S(t_0, \omega_0) |^2 \quad (5)$$

Figures 4(a) and 4(b) present this analysis for two representative phases, $\Delta\varphi$=0.7π and $\Delta\varphi$=π, respectively, revealing the sub-cycle electron dynamics.

Several key features emerge. First, for both cases, higher-order harmonics are emitted in distinct bursts localized near the extrema of the driving electric field, with the harmonic cutoff energy directly scaling with the peak field amplitude.[2, 49] This observation reinforces our conclusion that $\Delta\varphi$ controls the HHG spectrum by tailoring the transient field strength. Second, the sub-cycle waveform control offered by the two-color field directly sculpts the structure of the harmonic spectrum. For $\Delta\varphi$=0.7π, the field waveform features two adjacent peaks of comparable amplitude within a half-cycle [$A_1$ and $B_1$ in Figure 4(a)]. The emissions from these two events interfere, resulting in a harmonic spectrum with a sharply resolved, discrete structure.[50, 51] Conversely, for $\Delta\varphi$=π, the second harmonic constructively interferes to create a single, much stronger field maximum [$A_2$ in Figure 4(b)] and weaker secondary features ($B_2$, $C_2$). This produces a single, dominant emission burst per half-cycle, leading to a smoother, more continuous harmonic spectrum.

Crucially, in both scenarios, the primary harmonic emission events are locked to the extrema of the laser's vector potential. This specific timing is a definitive signature of HHG governed by interband dynamics, where electron-hole pairs are first created near the field zero-crossing, accelerated by the field, and then driven to recombine and emit high-energy photons as the vector potential reaches its maximum or minimum.[13, 28] This time-frequency analysis provides a compelling microscopic picture that fully supports our central conclusion - the HHG process in 1L-$WS_2$ is dominated by interband transitions, which can be precisely controlled on a sub-femtosecond timescale by the relative phase of a two-color field.

To elucidate the physical mechanism of the phase modulation effect, we analyze the dynamics of electronic excitation. Within the semiclassical acceleration theory, the motion of a Bloch electron under the laser pulse is governed by the time-dependent wave vector, $\boldsymbol{k}(t)=\boldsymbol{k}_0-e[\boldsymbol{A}(t)-\boldsymbol{A}(t_0)]/\hbar$, where $\boldsymbol{k}_0$ is the initial wave vector at the electron's birth time $t_0$, and $\boldsymbol{A}(t)$ is the vector potential of the two-color field.[52, 53] While TDDFT encapsulates the full spectrum of complex multi-band interferences, interpreting these dynamics through the three-step macroscopic perspective, as effectively applied in recent solid-state studies[31, 32], remains instructive.

Within this framework, the population of excited electrons serves primarily as a proxy for the initial step of the cycle - the interband transition (ionization) probability. As demonstrated in Figures 4(c) and 4(d), this initial excitation closely tracks the temporal structure of the vector potential modulus, $|\boldsymbol{A}(t)|$. However, the ultimate harmonic yield is inextricably linked not just to this initial population size, but to the subsequent coherent acceleration and recombination phases, which we track via the time-resolved orbital occupations [Figures 4(e) and 4(f)]. The dynamics differ markedly with the relative phase, $\Delta\varphi$. For $\Delta\varphi=0.7\pi$, the $|\boldsymbol{A}(t)|$ waveform exhibits a multi-peaked structure that periodically drives electron excitation, leading to a correspondingly oscillatory rise in the excited electron population. In stark contrast, for $\Delta\varphi=\pi$, constructive interference between the two-color fields reshapes the waveform into a single, dominant primary peak with significantly weaker sub-cycle features. This concentrates the excitation process into a much narrower temporal window, resulting in a single, prominent peak in the excited electron population. These results demonstrate that the sub-cycle structure of the vector potential, precisely controlled by $\Delta\varphi$, directly governs the temporal profile of electron excitation.

This phase-dependent excitation dynamic is mirrored in the time-resolved occupation of individual electronic orbitals [Figures 4(e) and 4(f)]. At $\Delta\varphi=0.7\pi$, the occupation numbers of multiple orbitals - spanning from VBM-5 to the CBM+3 - exhibit pronounced, periodic oscillations. This behavior is characteristic of coherent, Rabi-like cycling between the valence and conduction bands, driven by the periodic strong field.[54] This sustained quantum coherence ensures that recolliding electrons encounter a hole state with a well-defined phase relationship, a condition that is critical for efficient constructive interference and the emission of high-frequency harmonics.

Conversely, at $\Delta\varphi=\pi$, the orbital dynamics are governed by the impulsive nature of the single-peak vector potential. The occupations exhibit a stepped increase followed by rapid relaxation, distinct from the sustained, large-amplitude Rabi-like cycles observed at $\Delta\varphi$=0.7π [Figures 4(e) and 4(f)]. This rapid relaxation behavior of the occupancy number can be phenomenologically interpreted as destructive interference. Its origin can be attributed to the complex coupling between electronic states induced by the laser field at this specific phase. This causes the phases of different KS orbitals to rapidly lose synchronization during evolution, enabling efficient redistribution of the initial excitation energy among multiple electronic states. Consequently, the system evolves toward a hybrid-like state. The probability of coherent recombination between returning electrons and such desynchronized hole systems is significantly reduced, ultimately leading to strong suppression of the high-harmonic generation yield.

These findings reveal a clear microscopic picture - $\Delta\varphi$=0.7π enhances high-harmonic generation by maintaining quantum coherence in electron-hole dynamics, whereas $\Delta\varphi$=π suppresses harmonic generation by inducing strong non-adiabatic coupling between electronic states, leading to rapid, destructive interference behavior. Our detailed calculations of electronic transitions, as presented in Figures S1-S4, provide further insight into these mechanisms. Importantly, this not only confirms the dominant role of interband currents in $WS_2$ high-harmonic generation process at the microscopic level but also demonstrates that the relative phase of dual-color light serves as a powerful tool for regulating quantum coherent dynamics on ultrafast timescales.

Beyond optical control, we investigate strain engineering as an orthogonal means to tune the HHG process.[25, 55-58] As shown in Figure 5(a), applying biaxial strain has a pronounced effect - tensile strain systematically enhances the HHG yield, while compressive strain suppresses it. A polarization-resolved analysis reveals this effect to be highly anisotropic and provides a clear signature of underlying quantum geometric effects. While compressive strain (-2%) attenuates both polarization components, it particularly quenches the harmonic signal perpendicular to the strain axis ($P_y$) [Figure 5(b)]. Conversely, tensile strain (+2%) dramatically amplifies the $P_y$ component [Figure 5(c)]. This trend is quantified in Figure 5(d), which shows that tensile strain systematically increases the relative contribution of $P_y$ to the total harmonic power, demonstrating that strain anisotropically reshapes the pathways for harmonic emission.[58]

Our selection of three strain points (-2%, 0%, +2%) for the HHG simulations is grounded in both experimental feasibility and electronic topology robustness - biaxial strain up to ±2% has been reliably achieved in $WS_2$ monolayers via flexible substrates and nanobubble engineering,[57, 59, 60] and prior benchmark studies confirm that the LDA, while acknowledging its tendency to underestimate absolute band gaps, faithfully reproduces the strain-driven direct-to-indirect band gap transition topology in TMDs.[61] Our calculations of the system's electronic structure (covering -4%, -2%, 0%, +2%, +4%, see Figure 6) indicate that the most significant electronic topological change - namely, the transition from direct to indirect band gap - is most pronounced within the -2% compressive strain range. Nevertheless, we acknowledge that quantitative band-edge energies may shift with higher-level methods (e.g., GW approximation), and extreme strains (>±3%) lie beyond current routine experimental reach. Consequently, investigating this specific parameter space - spanning from the indirect-gap phase at -2% compressive strain to the direct-gap phases at 0% and +2% tensile strain - enables us to robustly capture the pronounced modulations in the HHG response governed by this underlying electronic topological transition.

Our analysis reveals that the strain dependence is governed by two distinct physical mechanisms. (1) Under compressive strain (-2%), the strong suppression of HHG is attributed to the direct-to-indirect bandgap transition [Figure 6(a)]. This topological change significantly modifies the electronic wavefunctions at the band edges. Crucially, it introduces a substantial momentum mismatch between the valence band maximum (VBM) and the conduction band minimum (CBM). In our rt-TDDFT simulations, which treat the electron dynamics within the frozen-ion approximation, this momentum mismatch directly and drastically reduces the interband transition dipole matrix elements near the original K-valleys [Supplementary Figure S11]. Consequently, the efficiency of laser-driven coherent electron excitation from the valence to the conduction band is severely diminished. Without efficient initial electron-hole pair generation, the subsequent steps of acceleration, recollision, and high-harmonic emission are effectively quenched on the sub-femtosecond timescale. (2) Under tensile strain (+2%), the amplification is driven by a profound reshaping of the BC [Figure 6(b)]. While the reduction of the direct bandgap enhances the overall carrier injection rate, the specific, dramatic boost in the perpendicular harmonic yield ($P_y$) originates from the strain-induced modification of the Bloch wavefunction geometry. As quantified in Table I, this geometric reshaping directly amplifies the anomalous velocity contribution to the emission current.

It is instructive to contrast our findings with prior work on $MoS_2$.[8] Guan *et al.* observed that compressive strain enhances HHG in $MoS_2$ by flattening the bands to boost intraband currents. In stark contrast, our analysis of $WS_2$ reveals that compressive strain suppresses HHG by inducing a direct-to-indirect bandgap transition [Figure 6(a)], which quenches the interband recombination process that dominates the emission in this material (Figure 2). Instead, we find that tensile strain enhances the yield. This comparison highlights that the nonlinear optical response of TMDs is highly material-specific, governed by the delicate competition between band topology and carrier coherence.

More profoundly, the anisotropic enhancement of the $P_y$ component is a direct manifestation of strain-induced changes in the material's quantum geometry. The BC, shown in Figure 6(b), acts as an intrinsic, momentum-space pseudo-magnetic field that gives rise to an anomalous velocity component perpendicular to the applied electric field.[62] In monolayer $WS_2$, strong spin-orbit coupling and broken inversion symmetry endow the $K$ and K' valleys with large BC of opposite sign.[63] Our calculations show that tensile strain, while preserving the $C_3$ rotational symmetry, systematically modifies the electronic wavefunctions and the associated BC. As detailed in Table I, applying +2% tensile strain increases the integrated BC near the K-valley from 18.24 $Å^{-2}$ to 26.86 $Å^{-2}$. This quantitative increase directly correlates with the near-doubling of the perpendicular harmonic yield (from 1467.0 to 2894.6 a.u.), confirming the anomalous velocity as the primary driver of the emission.

This amplification is rigorously governed by the anomalous velocity contribution to the intraband current. Within the semiclassical approximation, the field-driven electron velocity is given by $\boldsymbol{v}(\boldsymbol{k}) = \nabla_{\boldsymbol{k}}\varepsilon(\boldsymbol{k}) - \dot{\boldsymbol{k}} \times \boldsymbol{\Omega}(\boldsymbol{k})$. By substituting the time-dependent momentum change induced by the laser field, $\dot{\boldsymbol{k}} = -e\boldsymbol{E}(t)$, the anomalous velocity term becomes

$$\boldsymbol{v}_{anom}(t) \approx e\boldsymbol{E}(t) \times \boldsymbol{\Omega}(\boldsymbol{k}) \tag{6}$$

Since the driving laser field is linearly polarized along the zigzag direction ($\boldsymbol{E} \parallel \hat{x}$) and the BC is directed out-of-plane ($\boldsymbol{\Omega} \parallel \hat{z}$), this cross-product generates a purely transverse current density $J_y(t) \propto |\boldsymbol{E}(t)| \cdot \Omega_z(\boldsymbol{k})$. Consequently, the strain-induced enhancement of the integrated BC (Table I) acts as a direct linear multiplier for the transverse current, providing a deterministic mechanism for the monotonic amplification of the perpendicular harmonic yield ($P_y$) observed in our simulations. This establishes the perpendicularly polarized harmonic signal as a sensitive, all-optical probe of strain-modulated quantum geometry in 2D materials.

To elucidate the mechanism connecting strain to the harmonic yield, we computationally investigate the electronic excitation dynamics. Figure 7 presents the time-evolution of the excited electron population under varying strain conditions. Specifically, tensile strain ($\varepsilon$=+2%) leads to a significantly larger excited electron population compared to both the unstrained ($\varepsilon$=0%) and compressively strained ($\varepsilon$=−2%) cases [Figure 7(a)]. Conversely, compressive strain markedly suppresses the excitation efficiency. An analysis of the time-dependent orbital occupations reveals the origin of this behavior [Figures 7(b)-(d)]. In the unstrained case [Figure 7(b)], occupations near the VBM and CBM evolve smoothly, indicating that transitions are predominantly localized between these frontier bands. Under compressive strain [Figure 7(c)], oscillations in the VBM occupation are damped, and relaxation is accelerated. This is consistent with the strain-induced shift to an indirect bandgap, which suppresses efficient interband transitions. In sharp contrast, under tensile strain [Figure 7(d)], the populations of multiple conduction band orbitals (e.g., CBM, CBM+1) increase substantially and persist long after the laser pulse has passed. This sustained population of excited states signifies that tensile strain not only boosts excitation efficiency but also slows the subsequent carrier relaxation. This prolonged lifetime of excited carriers is crucial, as it increases the probability of their recollision with the parent ion - the final, decisive step in high-harmonic generation.

HHG, an advanced nonlinear optical phenomenon, demonstrates immense potential for generating coherent extreme ultraviolet (EUV) radiation and attosecond pulses. This study demonstrates that HHG yields in monolayer $WS_2$ can be effectively modulated by synergistically combining strain engineering with all-optical phase control, opening a new paradigm for coherent control in ultrafast optics and condensed matter physics. To systematically investigate the governing control mechanisms, we constructed a two-dimensional map of the HHG yield as a function of the two-color relative phase ($\Delta\varphi$) and applied strain ($\varepsilon$) [Figure 8]. Here, $\Delta\varphi$ provides all-optical control over the driving field, while $\varepsilon$ acts as a mechanical parameter that modifies the material's fundamental electronic and quantum geometric properties. In this map, the horizontal axis represents $\Delta\varphi$ (in units of $\pi$), and the vertical axis denotes $\varepsilon$ (ranging from -2% compressive to +2% tensile). The color scale (yellow to purple) indicates increasing HHG yield. This analysis reveals the independent contributions of strain and phase, as well as their synergistic interaction. First, strain engineering provides an effective knob for modulating the overall yield. While the total harmonic yield shows a moderate modulation of approximately 27.6% across the studied strain range (dominated by the robust parallel component), the perpendicular component - which encodes the quantum geometric information - exhibits a dramatic sensitivity, nearly doubling in intensity under tensile strain (see Table I). Compressive strain, therefore, consistently suppresses the HHG output. This modulation is attributed to strain-induced modifications of the electronic band structure and the associated BC, which directly influence the material's nonlinear optical response. Second, the efficacy of this all-optical phase control is strongly dependent on the strain state. Under tensile strain ($\varepsilon$=+2%), the HHG yield shows high sensitivity to $\Delta\varphi$, peaking at $0.3\pi$ and remaining elevated in the $0.3\pi$–$0.7\pi$ range. This indicates that fine-tuned regulation and maximization of HHG are possible via phase control, but only when the material is under optimal tensile stress. Conversely, under unstrained or compressive conditions, the influence of $\Delta\varphi$ on the yield diminishes significantly, resulting in a much flatter modulation profile. This synergistic effect reveals a crucial physical interplay - the two-color phase ($\Delta\varphi$) provides sub-femtosecond control over the electron-hole trajectories to optimize their recollision dynamics. However, this trajectory control is only effective when the material's electronic properties - namely the band structure and BC - are first favorably tuned by tensile strain. In compressed or unstrained states, the intrinsic band topology or interband transition dipoles hinder this phase-controlled optimization, rendering the all-optical control ineffective.

Although established through first-principles simulations, our proposed dual-mode control strategy heavily relies on accessible experimental capabilities. Biaxial strains up to ±2% are routinely achieved in monolayer TMDs via flexible substrates or nanobubble engineering[57, 59]. Simultaneously, sub-femtosecond phase modulation of two-color fields has become a cornerstone of attosecond spectroscopy[9, 17]. Translating these theoretical predictions to laboratory settings, however, necessitates evaluating the impact of experimental non-idealities, primarily laser fluctuations and intrinsic crystalline defects.

Stochastic fluctuations in the laser intensity or the relative phase, $\Delta\varphi$, will inherently induce inhomogeneous broadening of the harmonic peaks and potentially diminish the contrast of the phase-dependent yield modulation. Nevertheless, the defining experimental signature - the macroscopic amplification of the perpendicularly polarized harmonics under tensile strain - is fundamentally governed by the deterministic reshaping of the momentum-space BC. In monolayer $WS_2$, the BC acts as an intrinsic pseudo-magnetic field predominantly concentrated at the K and K' valleys. Because the transverse anomalous current scales linearly with this globally integrated geometric phase, it provides a robust, monotonic response to strain that remains highly distinguishable from stochastic temporal noise.

Furthermore, while intrinsic point defects (e.g., sulfur vacancies) inevitable in realistic $WS_2$ samples introduce mid-gap localized states that can alter interband transition dipoles and potentially quench the absolute HHG recombination efficiency, they primarily act as localized scattering centers. Conversely, the orthogonal control mechanism identified here originates from the global crystallographic symmetry and the intrinsic quantum geometric properties of the Bloch wavefunctions across the Brillouin zone. Such macroscopic topological features are inherently resilient to moderate local structural perturbations. Consequently, the fundamental physics of strain-amplified anomalous velocity will remain strictly observable, cementing the perpendicular harmonic yield as a robust optical probe of quantum geometry. Future experiments employing samples with controlled defect densities will be essential to quantitatively map the boundaries of this topological resilience.

## 4. Conclusions

In this study, we have employed first-principles rt-TDDFT to systematically investigate HHG and the underlying electronic dynamics in monolayer $WS_2$ under the combined influence of a two-color laser field and mechanical strain. Our results demonstrate that the relative phase ($\Delta\varphi$) of the two-color field provides a precise, sub-femtosecond control knob for manipulating the harmonic yield, spectral features, and polarization. Specifically, a relative phase of $\Delta\varphi$=0.7π optimizes the harmonic emission efficiency, a condition that correlates with sustained coherence in the electronic transitions. Our analysis further confirms that interband transitions are the dominant HHG mechanism, accounting for over 90% of the total harmonic intensity.

Furthermore, we establish that strain engineering offers a powerful and complementary method for tuning the material's nonlinear optical response. Applying tensile strain enhances the HHG efficiency and significantly amplifies the perpendicularly polarized harmonic component by strategically modifying the band structure and BC. Conversely, compressive strain suppresses harmonic emission; this is attributed to an increase in the bandgap and a shift toward an indirect-gap character, which collectively inhibit the requisite electronic excitations. Critically, the synergistic interaction between tensile strain and a specific relative phase difference maximizes the HHG yield. This synergistic effect is rooted in distinct physical mechanisms. strain-induced lattice distortion directly alters the material's band structure and quantum geometric properties, while the relative phase enables sub-femtosecond control of the electronic trajectories, thereby maximizing constructive interference.

Collectively, these findings elucidate the quantum dynamical pathways of solid-state HHG in monolayer $WS_2$ and provide a robust theoretical framework for designing efficient and highly tunable solid-state EUV light sources. By combining the distinct and complementary mechanisms of all-optical control and strain engineering, this work opens promising avenues for advanced applications in ultrafast spectroscopy and attosecond science. We note that the feasibility and robustness of the proposed mechanisms under realistic experimental conditions - including laser noise and potential material defects - warrant further investigation. While the topological nature of the BC provides inherent resilience against moderate local perturbations, subsequent studies incorporating explicit disorder, defect, and noise models will be crucial for translating this theoretical framework into practical devices.

**Declaration of competing interest**

The authors declare that they have no known competing financial interests or personal relationships that could have appeared to influence the work reported in this paper.

## ACKNOWLEDGMENTS

This work is sponsored by the Natural Science Foundation of Xinjiang Uygur Autonomous Region (Grant No. 2023D01D03 and 2022D01C689), the Xinjiang University Outstanding Graduate Student Innovation Project (Grant No. XJDX2025YJS154).

## Appendix A. Supplementary data

Supplementary data associated with this article can be found in the online version.

## Data availability

Data will be made available on request.

**Figures:**

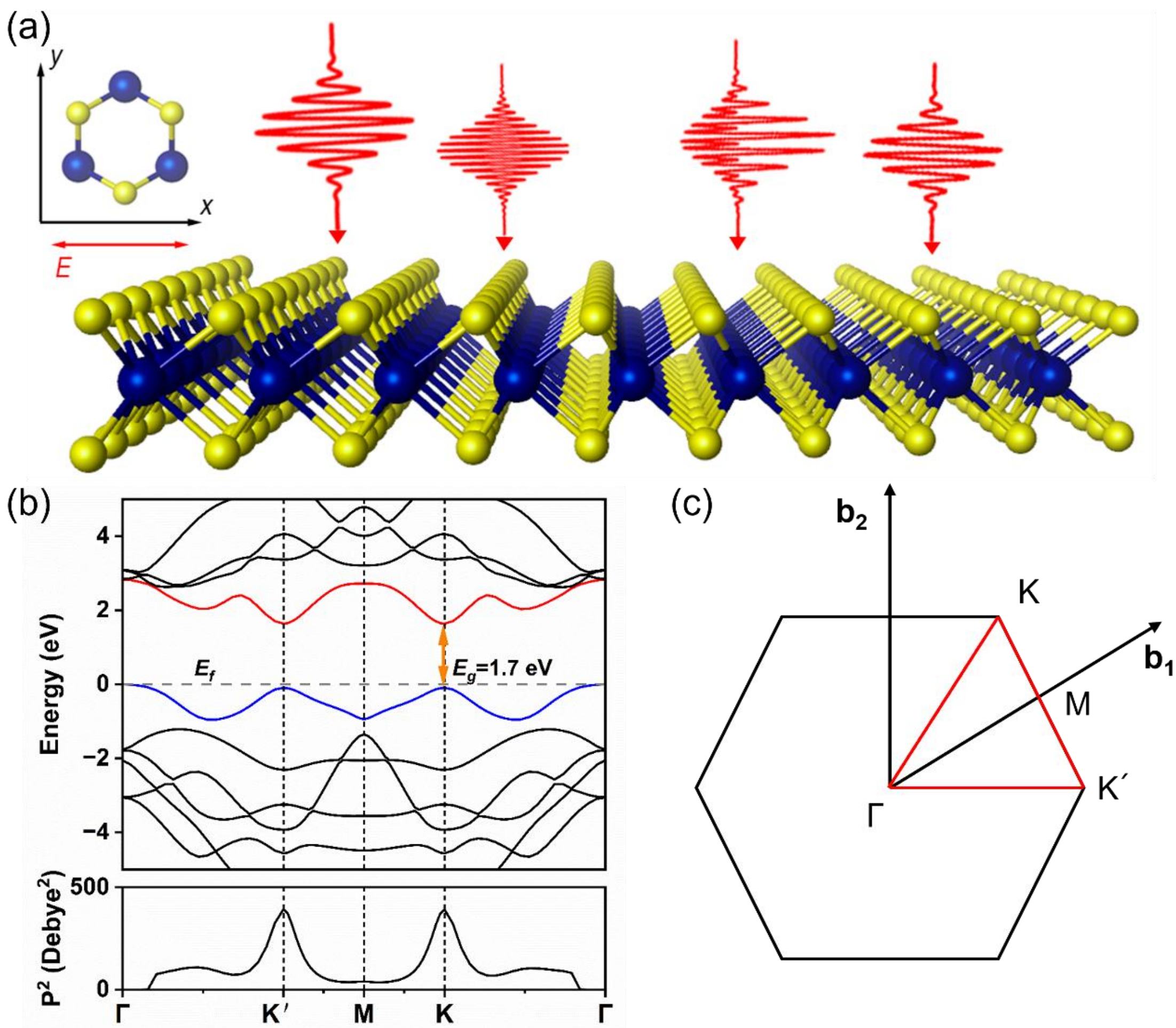

**Figure 1.** (a) Schematic of a monolayer (1L) $WS_2$ interacting with a two-color laser field. The laser propagates under normal incidence (along the *z*-axis), ensuring the electric field (***E***) is polarized strictly along the *x*-axis (zigzag direction). (b) Electronic band structure (top) and the corresponding squared transition dipole moment ($P^2$) (bottom) for 1L-$WS_2$. The horizontal dashed line at 0 eV indicates the Fermi level ($E_F$). The valence band maximum (VBM, blue) and conduction band minimum (CBM, red) are highlighted, showing a direct band gap ($E_g$) of approximately 1.7 eV at the K and K*'* valleys. (c) The first Brillouin zone of $WS_2$, showing the high-symmetry points Γ, M, K, and K', along with the reciprocal lattice vectors $\boldsymbol{b}_1$ and $\boldsymbol{b}_2$.

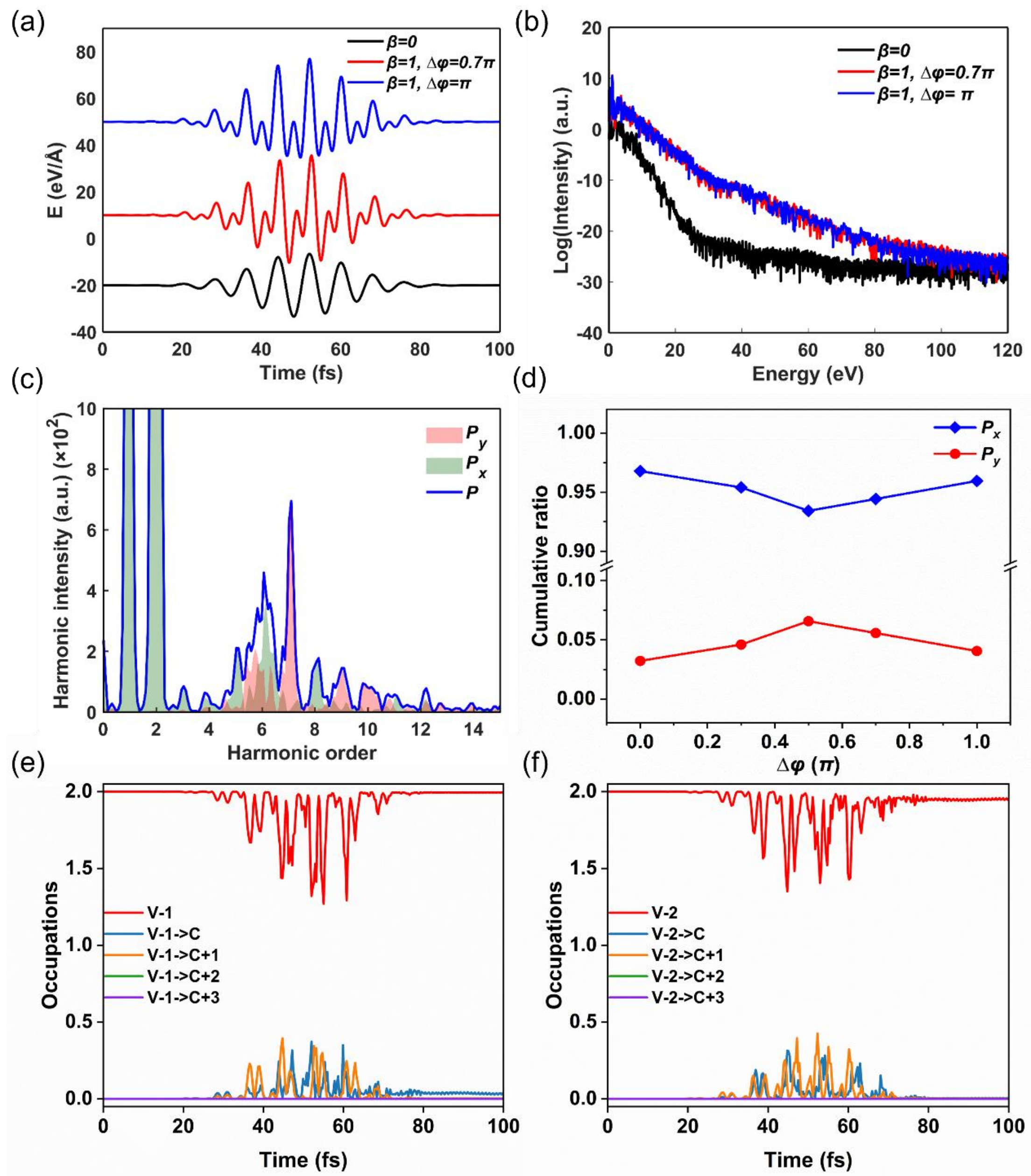

**Figure 2.** Control of high-harmonic generation (HHG) using a two-color laser field. The field consists of a fundamental frequency and its second harmonic, with a relative phase $\Delta\varphi$ and field strength ratio $\beta$. (a) Temporal profiles of the driving laser field for a single-color pulse ($\beta$=0) and two-color pulses with $\Delta\varphi$=0.7π and $\Delta\varphi$=π. Waveforms are vertically shifted for clarity. (b) Corresponding HHG spectra, plotted on a logarithmic scale. (c) Polarization-resolved harmonic spectrum ($P_x$ and $P_y$) for $\Delta\varphi$=0.7π. (d)

Fractional contribution of the parallel and perpendicular components to the total harmonic power for the five relative phases shown. (e-f) Contributions to high-harmonic generation (HHG) from valence-to-conduction band transitions at $\Delta\varphi=0.7\pi$, illustrating the analogous contributions from transitions originating from the VBM-1 and VBM-2 bands, respectively.

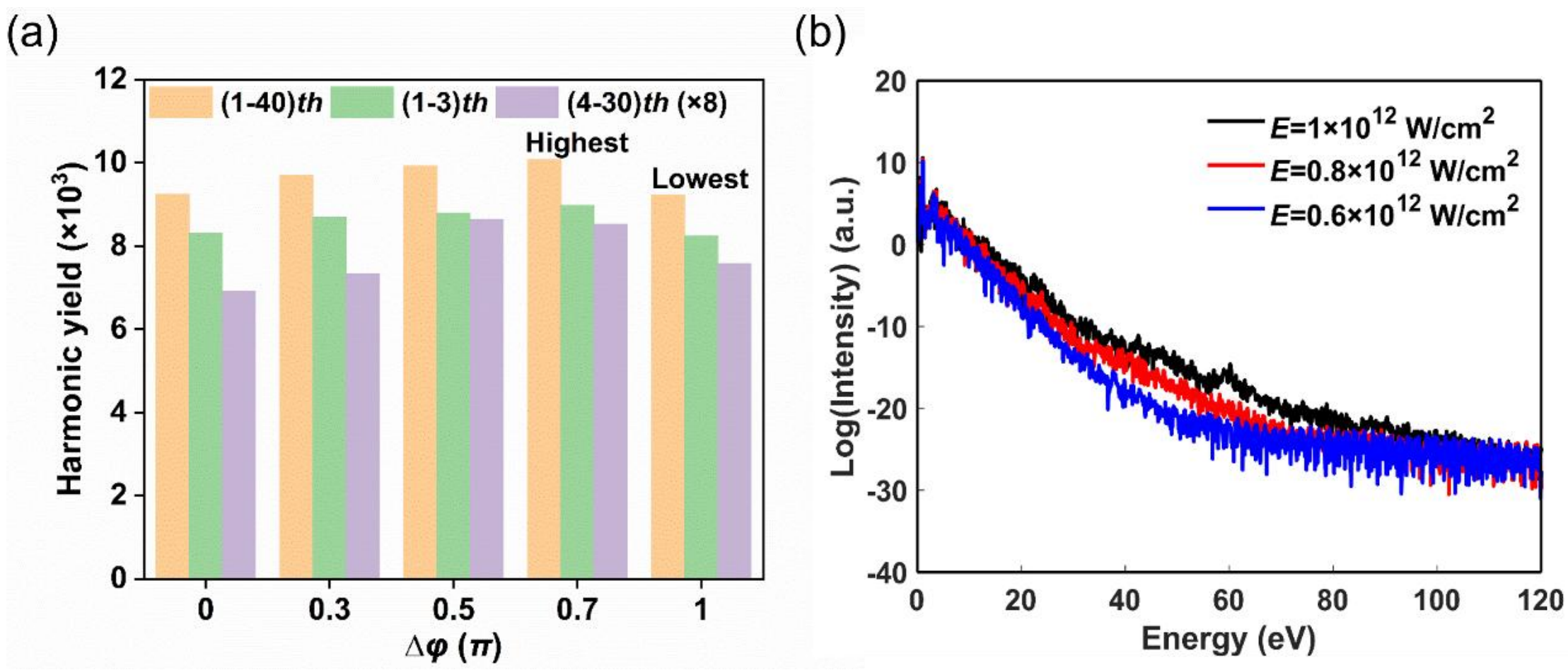

**Figure 3.** (a) Integrated harmonic yield as a function of the relative phase, $\Delta\varphi$, for the total spectrum (orders 1-40), low-order harmonics (1-3), and high-order harmonics (4-30). In (e), the high-order harmonic yield is scaled by a factor of 8 for clarity. (b) HHG spectra calculated at different laser intensities. The curves correspond to intensities of $E$=1×10$^{12}$ W/cm$^2$ (black), 0.8×10$^{12}$ W/cm$^2$ (red), and 0.6×10$^{12}$ W/cm$^2$ (blue), showing the dependence of harmonic yield on the field strength.

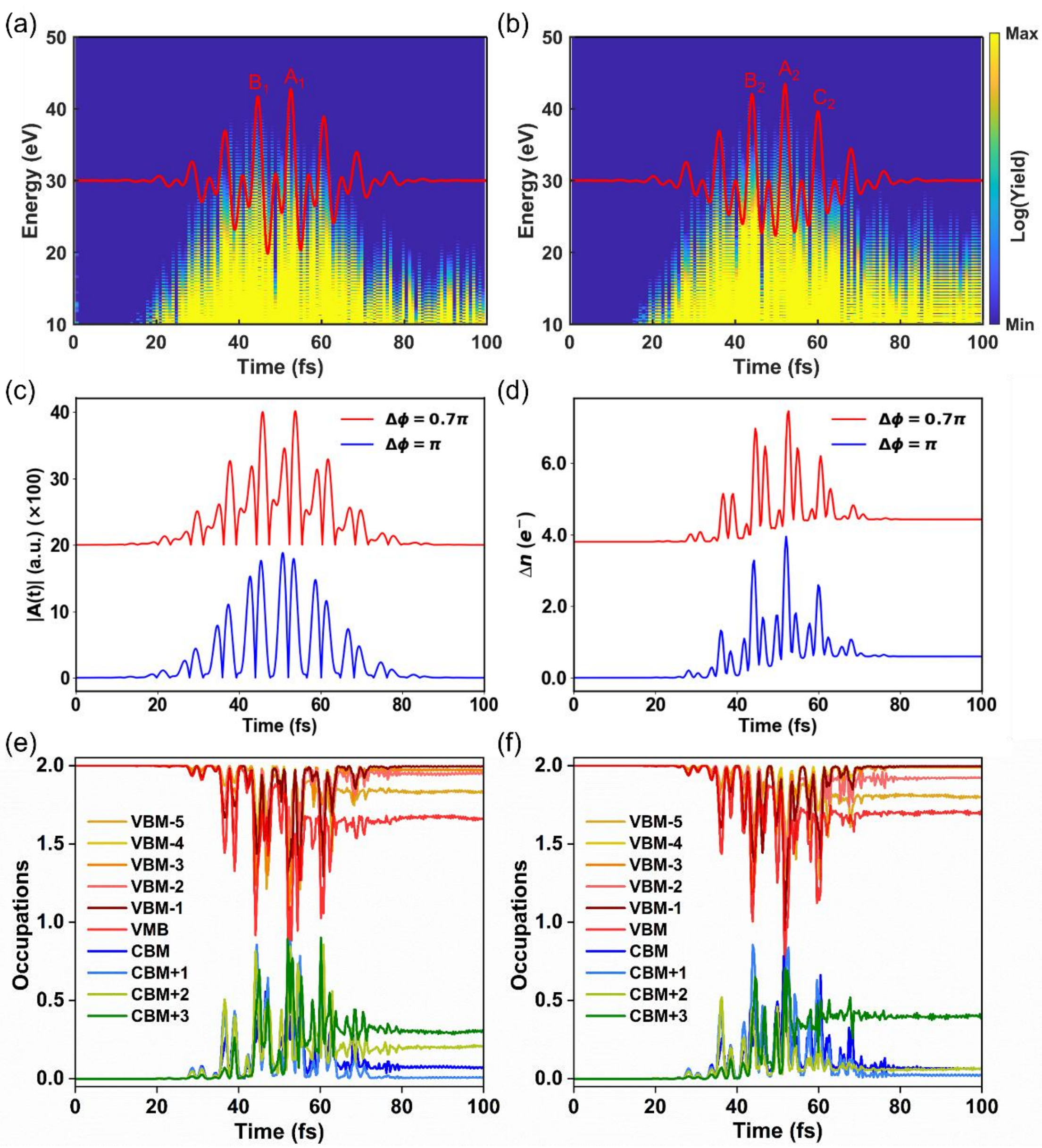

**Figure 4.** Time-resolved electron dynamics under two-color fields. (a, b) Time-frequency analysis of the harmonic emission for $\Delta\varphi$=0.7π (a) and $\Delta\varphi$=π (b). The overlaid red curves show the driving electric field, and the labels (e.g., $A_1$, $B_1$) denote distinct emission bursts. Unless otherwise stated, the laser intensity is $1.0\times10^{12}$ W/cm$^2$ and $\beta$=1. (c) Temporal profile of the vector potential amplitude, $|\boldsymbol{A}(t)|$, and (d) time evolution of the total excited electron population, $\Delta n$, for relative phases of $\Delta\varphi$=0.7π (red) and $\Delta\varphi$=π (blue). For visibility, the $\Delta\varphi$=0.7π curves in (c) and (d) are vertically offset by +20 and +3.8, respectively. (e, f) Time-dependent occupations of specific valence (VBM) and conduction (CBM) bands for (e) $\Delta\varphi$=0.7π and (f) $\Delta\varphi$=π.

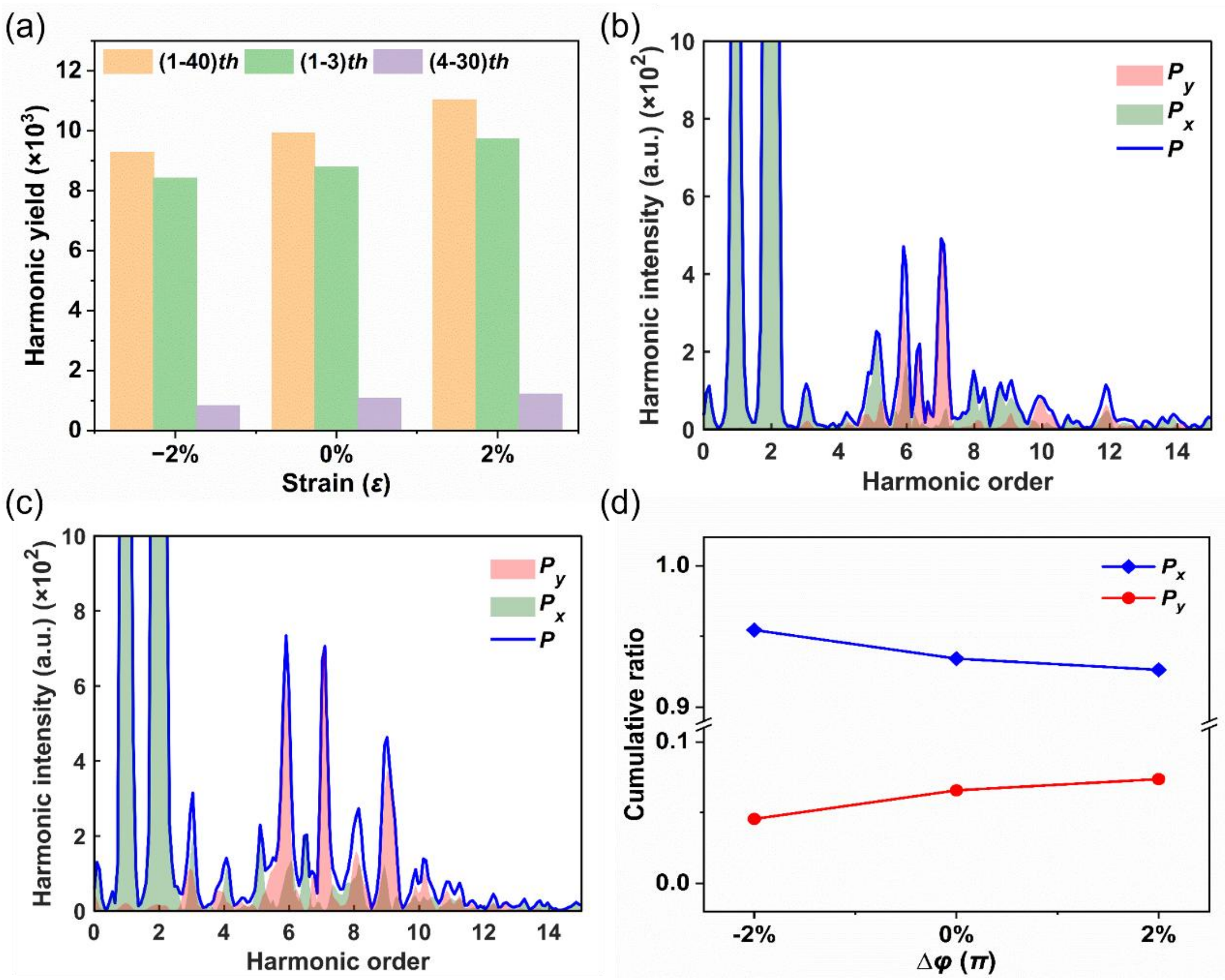

**Figure 5.** Effect of biaxial strain ($\varepsilon$) on high-harmonic generation. (a) Integrated harmonic yield as a function of strain for the total spectrum [(1-40)th], low-order harmonics [(1-3)th], and high-order harmonics [(4-30)th]. The yield values are scaled by $10^3$. (b, c) Polarization-resolved HHG spectra under (b) compressive strain (−2%) and (c) tensile strain (+2%). (d) Fractional power distribution between the harmonic components parallel ($P_x$) and perpendicular ($P_y$) to the driving laser's polarization.

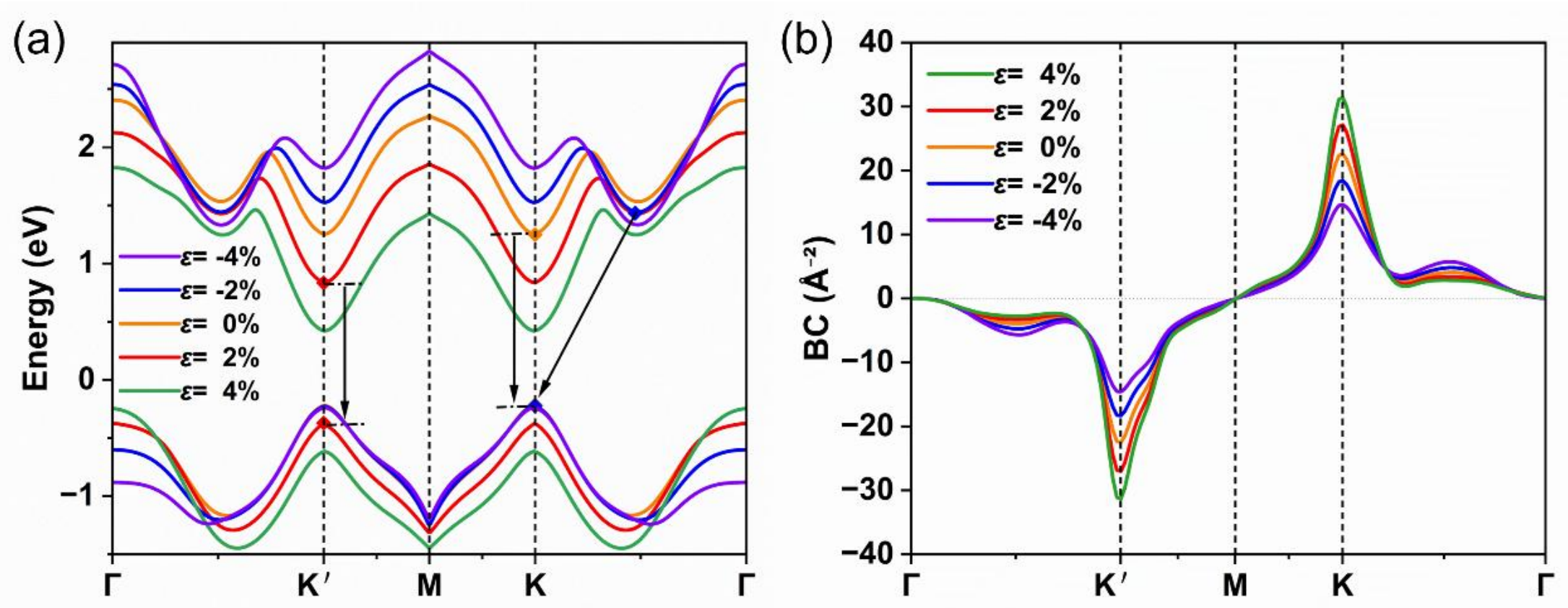

**Figure 6.** Effect of biaxial strain (ε) on the electronic properties of monolayer $WS_2$. (a) Electronic band structure and (b) the corresponding Berry curvature (BC) calculated along the high-symmetry path Γ-K′-M-K-Γ. The calculations are shown for strain values ranging from -4% (compressive) to +4% (tensile). The direct and indirect band gaps are indicated by black arrows in subfigure (a).

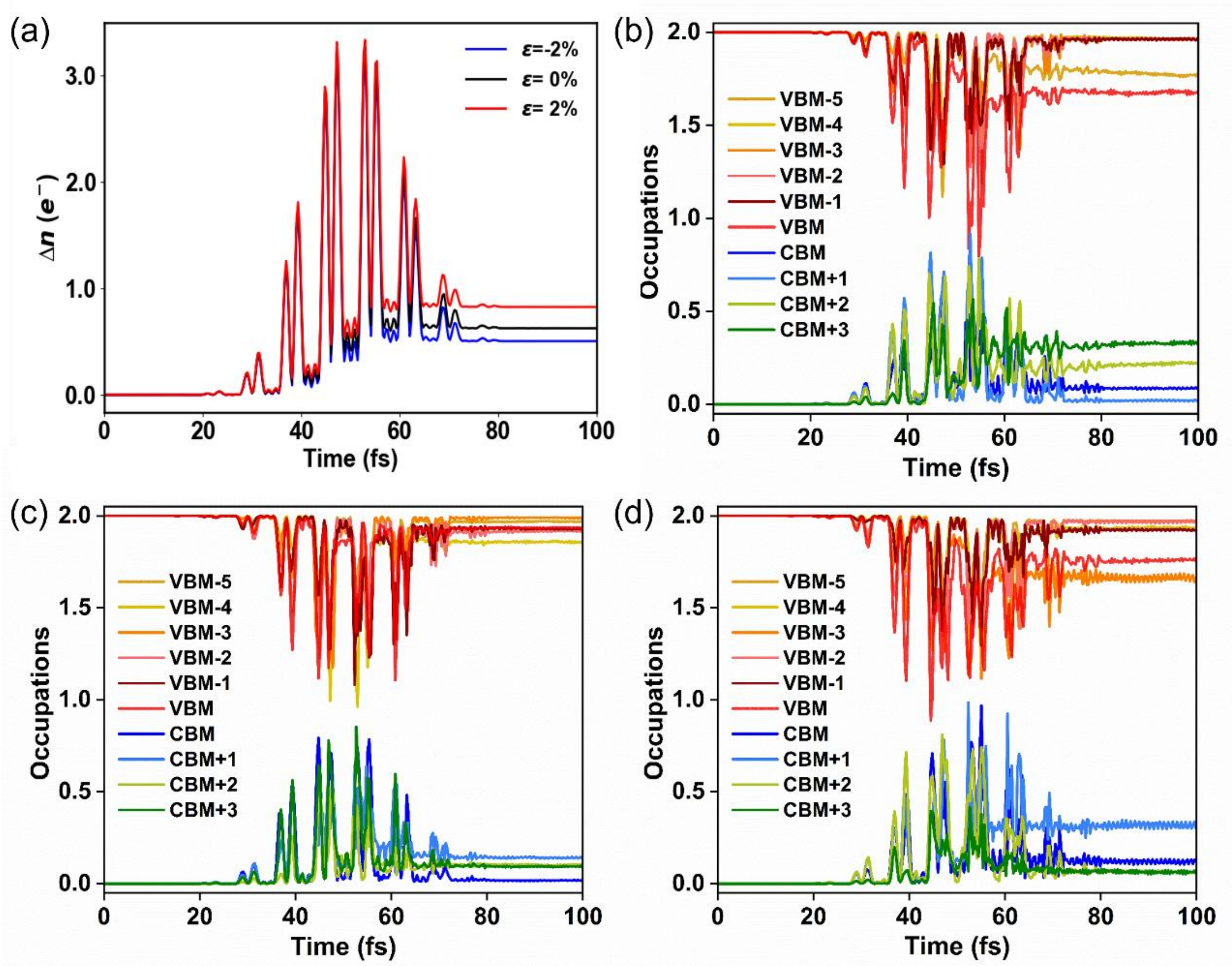

**Figure 7.** Effect of biaxial strain on time-resolved electron dynamics. (a) Time evolution of the total excited electron population ($\Delta n$) for compressive (−2%), zero (0), and tensile (+2%) strain. (b-d) The corresponding state-resolved occupations in the valence (VBM) and conduction (CBM) bands under conditions of (b) zero strain, (c) −2% compressive strain, and (d) +2% tensile strain.

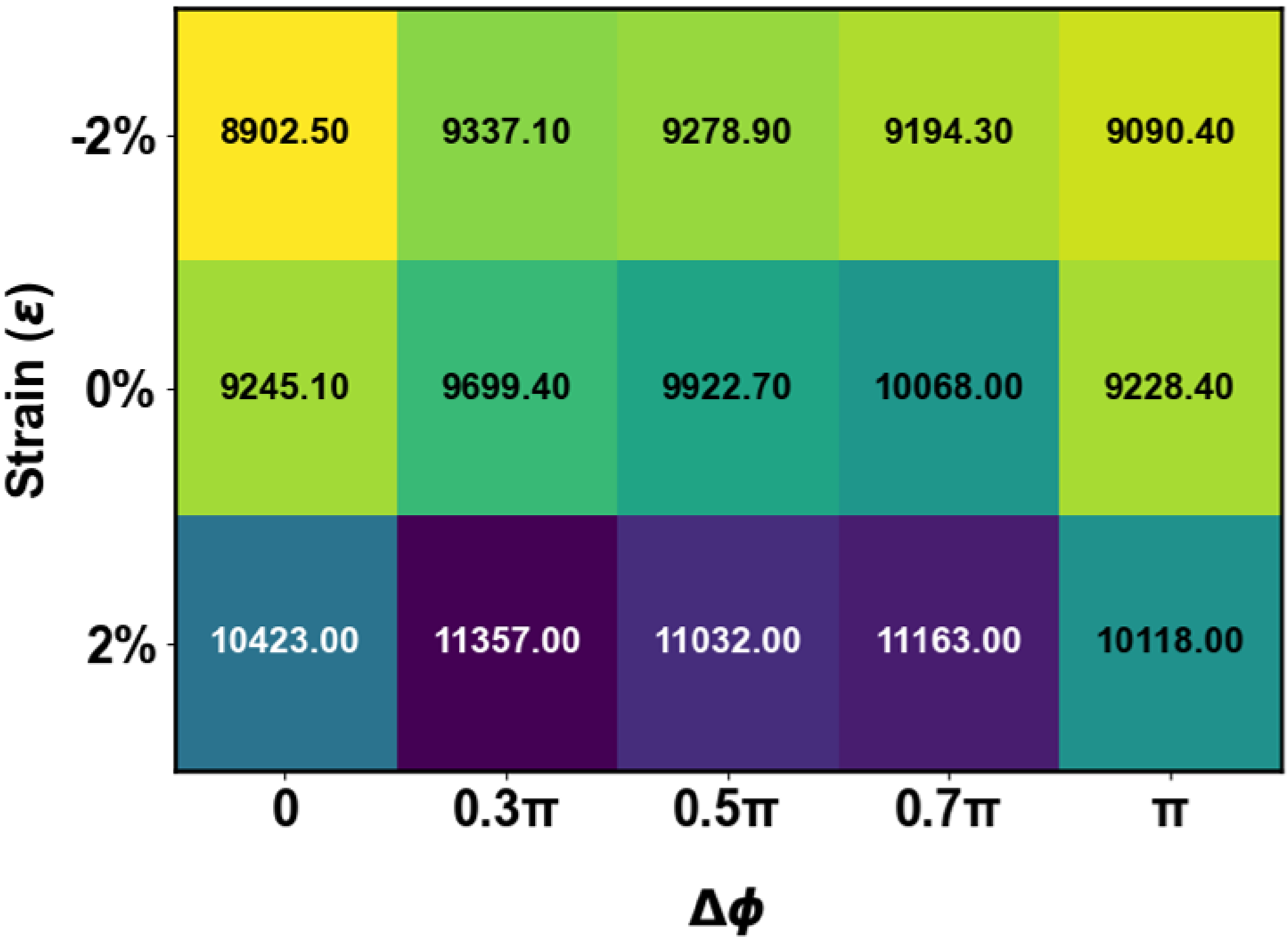

**Figure 8.** Synergistic regulation of harmonic anisotropy. The 2D heatmap plots the harmonic anisotropy as a function of biaxial strain ($\varepsilon$) ($y$-axis) and the relitive phase difference ($\Delta\varphi$) ($x$-axis).

**Tables:**

**TABLE I.** Comparison of Berry curvature (BC) and harmonic yield. The table lists three distinct values for BC (in $Å^{-2}$) and their corresponding harmonic yields (in a.u.).

| **BC ($Å^{-2}$)** | **Harmonic yield (a.u.)** |
|---|---|
| 18.24 | 1467.0 |
| 22.36 | 2285.1 |
| 26.86 | 2894.6 |